\documentclass[aps,twocolumn,groupedaddress]{revtex4}

\usepackage{epsfig}
\usepackage[fleqn]{amsmath}
\usepackage{amsfonts}
\usepackage{amssymb}
\usepackage{psfrag}
\usepackage{hvdashln}

\newcommand{\eq}[1]{\begin{equation} #1 \end{equation}}

\newcommand{\al}{\alpha}
\newcommand{\be}{\beta}
\newcommand{\sig}{\sigma}

\newcommand{\moy}[1]{\left\langle#1\right\rangle}

\newcommand{\eps}{\epsilon}

\newcommand{\vect}[1]{\boldsymbol{\rm #1}}
\newcommand{\ver}{\vect{r}}
\newcommand{\vmu}{\vect{\mu}}

\newcommand{\rom}[1]{\mathrm{#1}}
\newcommand{\romd}{\rom{d}}

\newcommand{\grad}{\vect{\nabla}}

\begin{document}



\title{ Structure and dielectric properties of polar fluids with extended dipoles: \\ results from numerical simulations
}

\author{V. Ballenegger}
\email{vcb25@cam.ac.uk}
\author{J.-P. Hansen}

\affiliation{Department of Chemistry\\
University of Cambridge\\
Cambridge CB2 1EW (UK)
}

\date{\today}

\begin{abstract}
The strengths and short-comings of the point-dipole model for polar fluids of spherical molecules are illustrated by considering the physically more relevant case of extended dipoles formed by two opposite charges $\pm q$ separated by a distance~$d$ (dipole moment $\mu=q d$). Extensive Molecular Dynamics simulations on a high density dipolar fluid are used to analyse the dependence of the pair structure, dielectric constant~$\eps$ and dynamics as a function of the ratio $d/\sigma$ ($\sig$ is the molecular diameter), for a fixed dipole moment $\mu$. The point dipole model is found to agree well with the extended dipole model up to $d/\sig \simeq 0.3$. Beyond that ratio, $\eps$ shows a non-trivial variation with~$d/\sig$. When $d/\sig>0.6$, a transition is observed towards a hexagonal columnar phase; the corresponding value of the dipole moment, $\mu^2/\sig^3 k T=3$, is found to be substantially lower than the value of the point dipole required to drive a similar transition.
\end{abstract}

\maketitle


\section{Introduction}
\label{sec:introduction}

Highly polar fluids are particularly important in many areas of Physical Chemistry, Chemical Engineering and Biology, because of their role as solvents leading to electrolyte and polyelectrolyte dissociation. Water is of course the most important among polar liquids, but because of its complex behaviour, primarily linked to the formation of hydrogen-bond networks, much theoretical work has focussed on simpler models involving spherical molecules with point dipoles. The best-known and widely studied examples are dipolar hard spheres, and the Stockmayer model (dipolar + Lennard-Jones interactions). A long-standing problem, going back to the classic work of Onsager~\cite{Ons} and Kirkwood~\cite{Kir}, is to relate the dielectric response of a polar fluid to molecular dipole fluctuations and correlations~\cite{MadKiv}. Subtle conceptual and numerical problems arise in Molecular Dynamics or Monte Carlo simulations of finite samples of polar fluids, which are linked to the infinite range of the dipolar interactions, so that boundary conditions must be treated adequately. These issues were resolved in the early eighties, both for the reaction field and the Ewald summation implementations of boundary-conditions~\cite{PatLevWei,deLPerSmi,Neu}. Despite this theoretical progress, accurate estimates of the dielectric permittivity of simple polar fluids by numerical simulation remains a very challenging task, because large fluctuations of the total dipole moment of the sample occur on a relatively long timescale (of the order of 10\,ps), leading to a very slow convergence rate for the dielectric constant \cite{Kus90,KurMarCha} (see also Sct. III).

More recently, it was realized that simple dipolar liquids can form a ferroelectric nematic phase for sufficiently large dipole moments~\cite{WeiPat,WeiLev,GroDie}. This transition is intimately related to the formation of chains of dipoles aligned head-to-tail, which prevent the formation of a proper liquid phase in the Stockmayer model if the dispersive energy is below a certain threshold~\cite{Smi}.

However point dipoles represent a limiting situation, never achieved in real polar molecules, which are characterized by extended charge distributions linked to electronic charge transfer from electron donors to electron acceptor atoms. In simple heteronuclear diatomic molecules like CO or HF, this situation can be modelled by assigning fractional charges of opposite sign to sites which are separated by a distance $d$, typically of the order of 0.1\,nm \cite{KleMcD}. Such, or more complicated situations involving more than two atoms, can be mimicked by adding higher order point multipoles to a point dipole~\cite{dipqua}, but such an expansion will require more and more high order multipoles as two molecules approach each other.

In this paper, we present a systematic investigation of the structure, dielectric response and phase behaviour of a simple model involving spherical molecules carrying extended (rather than point) dipoles resulting from opposite charges $\pm q$, each displaced symmetrically by a distance $d/2$ from the centre of the molecule. We study how the properties of the polar liquid change when $d$ is increased from zero, varying $q$ simultaneously so that the dipole moment $|\vmu|=qd$ remains constant. Although polar molecules are never spherical, the model investigated in this paper, which focusses on the electrostatic rather than steric interactions, is the simplest ``natural'' extension of the dipolar sphere model towards a more realistic representation of highly polar fluids. Some studies on the structure of similar models with extended dipoles have been published previously, but without an investigation of their bulk dielectric properties \cite{Cic,DieMor}.

\section{The model and simulation details}
\label{S2}

\begin{figure}[h]
\begin{center}
\psfrag{d}{$\vect{d}$}
\psfrag{s}{$\sig$}
\epsfig{file=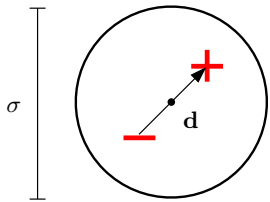}
\caption{\label{fig1} A polar molecule with an extended dipole moment}
\end{center}
\end{figure}

We consider a polar fluid made up of soft spheres with two embedded point charges $\pm q$ located at $\pm \vect{d}/2$ from the centre of the molecule (see Fig.~\ref{fig1}). The distance $|\vect{d}|$ is assumed fixed, so the molecule is not polarizable and carries a permanent dipole moment $\vect{\mu}=q \vect{d}$.

Placing the origin at the centre of the sphere, the multipole moments $q_{lm} = \int Y_{lm}^*(\theta,\phi) r^l\rho(\ver)\romd^3\ver$, where $\rho(\ver)=q\delta(\vect{d}/2)-q\delta(-\vect{d}/2)$ is the molecular charge distribution \cite{Jac}, are 
\eq{
  q_{lm}  = \begin{cases}
   2q \left( \frac{d}{2}\right)^{l} \sqrt{\frac{2l+1}{4\pi}} & \text{if $l$ odd and $m=0$}	\\
   0	& \text{otherwise.}	\\
   \end{cases}
}
The next non-vanishing moment after the dipole is thus the octopole, since the quadrupole moment vanishes by symmetry for this choice of origin.

\begin{figure}[h]
\begin{center}
\psfrag{beta_Uint}{$\beta U_{\rm int}$}
\psfrag{inset}{\epsfig{file=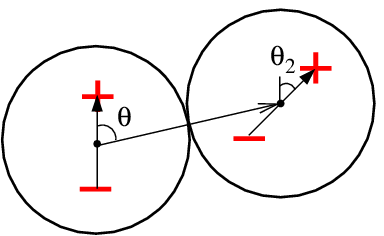,scale=0.5}}
\hspace{-3mm}
\epsfig{file=U_int_dipoct.eps,scale=0.37}
\caption{\label{fig2}Electrostatic interaction energy of two molecules at contact ($|\ver|=\sig$) for $d=\sig/2$.}
\end{center}
\end{figure}

The interaction energy between two molecules is given by the sum of a soft sphere repulsion $v_{ss}(\ver)=4u \sigma^{12}/|\ver|^{12}$ and the four Coulombic energies due to the point charges.
On Fig.~\ref{fig2}, the electrostatic energy at contact is compared to a truncated multipolar expansion containing the dipole-dipole and dipole-octopole interactions. The configuration of lowest energy occurs when the molecular dipoles are aligned head-to-tail ($\theta=\theta_2=0$). This minimum energy is lower for extended than for point dipoles.

The polar fluids we have considered can be characterized by the following dimensionless~\footnote{Our parameters in dimensioned units were $T=300\,$K, $\mu=2.45\,$D, $\sig=0.365502\,$nm, $m=10\,$u, $I=0.15625$\,u$\cdot$nm$^2$, $u=1.84767$\,kJ/mol.} parameters:

\begin{tabular}{ll}
$\ \ \bullet$ reduced density: & $\rho^* = \rho\sig^3 = 0.8$ \\
$\ \ \bullet$ reduced temperature: & $T^* = {kT}/{u} = 1.35$	\\
$\ \ \bullet$ reduced dipole moment: & $\mu^* = \sqrt{{\mu^2}/{\sigma^3 u}} = 2.0$ \\
$\ \ \bullet$ reduced moment of inertia: & $I^* = {I}/{m\sig_{\phantom{M}}^2}=0.117.$ \\
\end{tabular}

\noindent This corresponds to the same thermodynamic state point that has been extensively studied by Kusalik in the case of point dipoles~\cite{Kus90,Kus91}. See \cite{SteGre} for an investigation of the phase diagram of the dipolar soft-sphere system. Equilibrium quantities, such as the dielectric constant and distribution functions, are independent of the moment of inertia.

In all calculations, we employed periodic boundary conditions. We chose the spherical geometry, i.e. the periodic replications of the basic cubic simulation cell form an infinite sphere, which is itself embedded in a infinite region of dielectric constant $\eps'$. In this case, the Hamiltonian of the system is
\eq{	\label{H}
   H = \sum_{i<j=1}^N (v_{ss}(\ver_{ij})+q_i q_j \Psi(\ver_{ij})) + \frac{2\pi \vect{M}^2}{(2\eps'+1)L^3}
}
where $L$ is the side of the box, $\vect{M}=\sum_i q_i\ver_i$ is the total dipole moment, and
\begin{multline}	\label{def Psi}
   \Psi(\ver) = \sum_{\vect{n}\in\mathbb{Z}^3} \frac{\mathrm{erfc}(\kappa|\ver+\vect{n}L|)}{|\ver+\vect{n}L|} + \\ + \frac{1}{\pi L}\sum_{n\neq0}\frac{1}{n^2}\exp\left(\frac{-\pi^2n^2}{\kappa^2L^2}+\frac{2\pi i}{L}\vect{n}\cdot\ver\right).
\end{multline}
The last term in \eqref{H} accounts for the work done against the depolarizing field due to the surface charges induced on the spherical boundary. This term vanishes only for metallic boundary conditions ($\eps'=\infty$). The Ewald sums in $\Psi(\ver)$ were evaluated using the Smooth Particle Mesh Ewald method~\cite{PME} (Ewald coefficient $\kappa=3.4705$\,nm$^{-1}$, grid size 32x32x32, interpolation order~6). The interactions were truncated beyond 0.9\,nm, both for the real space Ewald sum and for the soft-sphere repulsions. 

Molecular Dynamics simulations were carried out using the simulation package {\it gromacs}~\cite{gromacs}. The equations of motion were integrated using the so-called leap-frog algorithm with a reduced time step of $\romd t^* = {dt}/{\sqrt{m \sig^2/u}} = 0.00235$. The temperature was kept constant using a Berendsen thermostat. Equilibration periods lasted at least 100\,ps ($\simeq$118 reduced time units), and were followed by data producing runs of 8\,ns or more. The number of molecules was 512 in calculations of the dielectric constant (Sct. III), and 5555 in calculations of correlation functions (Sct. IV).

\section{Dipole fluctuations and dielectric constant}
\label{S3}

The dielectric constant of a homogeneous and isotropic fluid can be calculated from the fluctuation formula~\cite{FreSmi}
\eq{	\label{KF}
   \frac{(\eps - 1)(2\eps'+1)}{2\eps'+\eps} = \frac{4\pi}{3V} \frac{\moy{\vect{M}^2}}{kT},
}
which holds for a macroscopic spherical sample of volume $V$ surrounded by a medium of dielectric constant~$\eps'$. The results obtained for the dielectric constant are independent of the choice of $\eps'$, provided the boundary term in eq.~\eqref{H} is properly taken into account. We employed metallic boundary conditions, because they are known to produce less uncertainties in estimates of $\eps$ than boundary conditions with finite $\eps'$ \cite{deLPerSmi,WanHolMul} (see also below). The fluctuation formula reduces in this case to
\eq{	\label{KF cond}
   \eps = 1 + 3y\moy{g}, \quad g=\frac{M^2}{N\mu^2},
}
where the dimensionless parameter $y=4\pi\be \rho \mu^2/9$ equals 3.309 at the state point under consideration.

\begin{figure}
\begin{center}
\psfrag{e}{\hspace{-10ex}Dielectric constant $\epsilon$}
\psfrag{d*}{\hspace{-10ex} Elongation $d/\sigma$}
\epsfig{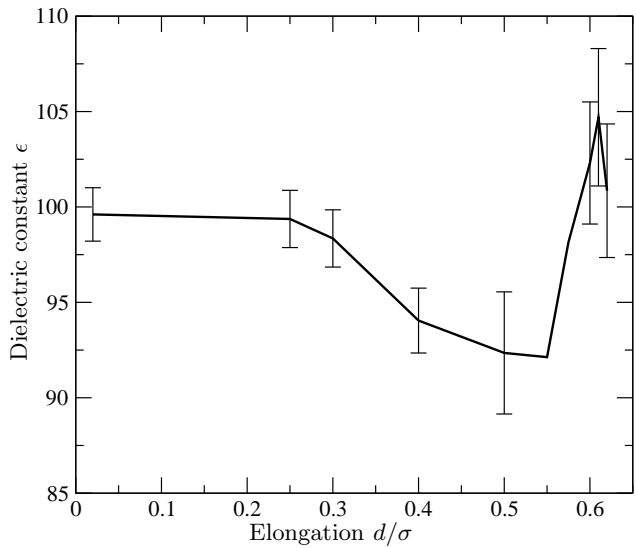}
\caption{\small\label{fig3}Dielectric constant of a dipolar soft sphere fluid as a function of dipole extension ($\rho^*=0.8$, $T^*=1.35$, $\mu^*=2$).}
\end{center}
\end{figure}

\begin{table}[h]
\begin{center}
\begin{tabular}{cccccc}
\hline
$d/\sig$	&	$\eps$	&	$\tau_M$ [ps]	&	$\tau_\mu$ [ps]	&	$D$ [$10^{-5}$\,cm$^2$/s]	& $p^*$\\	\hline
0.02	&	99.6($\pm$1.4)	&	2.21		&	0.50	&	$11.7$	&	0.50	\\
0.3	&	98.4($\pm$1.5)	&	2.64		&	0.54	&	$11.6$	&	0.53	\\
0.4	&	94.0($\pm$1.5)	&	2.88		&	0.63	&	$11.5$	&	0.53	\\
0.5	&	92.4($\pm$1.7)	&	4.14		&	0.88	&	$10.6$	&	0.49	\\
0.6	&	102.3($\pm$3.2)	&	11.44		&	2.01	&	$8.7$	&	0.31	\\
0.61	&	104.7($\pm$3.6)	&	13.97 	&	2.36	&	8.5	&	0.27	\\
0.62	&	100.8($\pm$3.5)	&	14.81		&	2.79	&	7.9	& 0.22		\\
\hline
\end{tabular}
\end{center}
\caption{\small\label{T1} Influence of dipole elongation on some properties of a polar soft sphere fluid.}
\end{table}

\begin{figure}[h]
\begin{center}
\psfrag{C_M}{{\footnotesize $C_{\vect{M}}(t)$}}
\psfrag{C_mu}{{\footnotesize $C_{\vect{\mu}}(t)$}}
\epsfig{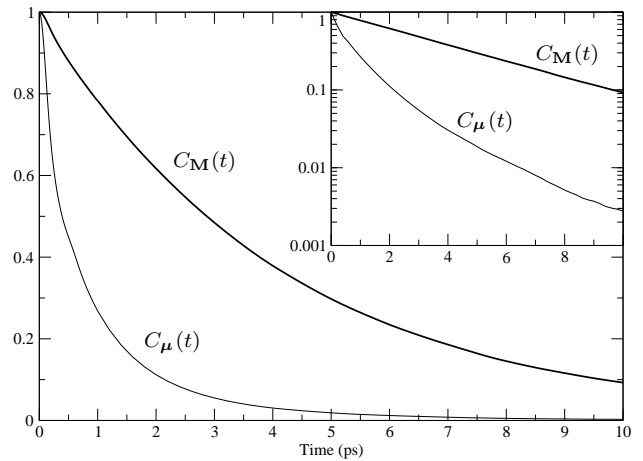}
\caption{\small\label{fig4}Autocorrelation functions $C_{\vect{M}}(t)$ and $C_{\vmu}(t)$, for the value $d=\sig/2$. The inset shows a logarithmic plot, confirming the exponential behaviour of $C_{\vect{M}}(t)$.}
\end{center}
\end{figure}

We show in Fig.~\ref{fig3} the variation of the dielectric constant with the extension $d$ of the dipole, at fixed dipole moment $\mu^*=2$. Table~\ref{T1} lists the actual values of~$\eps$, together with some other quantities of interest, namely the diffusion constant $D$, the dielectric relaxation times $\tau_{M}$ and $\tau_\mu$, and the reduced pressure $p^*=p\,\sig^3/u$.  The diffusion constant was calculated from Einstein's relation
\eq{
   \moy{|\ver_i(t)-\ver_i(0)|^2} = 6Dt, \qquad t\to\infty.
}
The relaxation times $\tau_M$ and $\tau_\mu$ were determined from the autocorrelation functions $C_{\vect{M}}(t) = \moy{\vect{M}(t)\cdot\vect{M}(0)}/$ $\moy{{M}^2}$ and $C_{\vect{\mu}}(t)$ (see Fig.~\ref{fig4}). For $t>0.3$\,ps, $C_{\vect{M}}(t)$ exhibits an exponential decay $\exp(-t/\tau_M)$ typical of a Debye dielectric. The relaxation of $C_{\vmu}(t)$ is not well fitted by a single exponential, and the corresponding relaxation time was estimated from the integral of  $C_{\vmu}(t)$.

For almost point dipoles ($d^*=d/\sig=0.02$), our value for the dielectric constant is in good agreement with the result $\eps_{\rm point}=98\pm2$ reported by Kusalik {\it et al.} for the ideal dipolar soft sphere fluid~\cite{KusManSvi}. Our data shows that when $d^*$ increases, the dielectric constant decreases and reaches a minimum about 6\% lower than $\eps_{\rm point}$ at $d^*\simeq 0.55$. When $d^*$ is further increased, the dielectric constant increases rapidly above $\eps_{\rm point}$, up to the critical distance $d_c^*\simeq 0.63$. At this critical distance, a phase transition occurs from an isotropic fluid to an orientationaly ordered ``liquid crystal'' phase (see Sct.\ref{S6}).

The simulations show that the point dipole model gives a reliable estimate of the dielectric constant over a very wide range of extensions $d$, namely up to the point where the system undergoes a phase transition. The weak sensitivity of the dielectric constant on the extension of the dipole, which contrasts with the large sensitivity observed in water models~\cite{HocBorBitSte}, may be due to the absence of a quadrupole moment in our molecules.

It is clear from Table~\ref{T1} that the dynamics of the fluid slows down when $d$ is increased: the diffusion coefficient $D$ drops and the relaxation times increase. This slowdown is due to the formation of head-to-tail dipolar chains in the system. Their entanglement make these chains less mobile than individual molecules in the present high density regime.

\begin{figure}[h]
\begin{center}
\epsfig{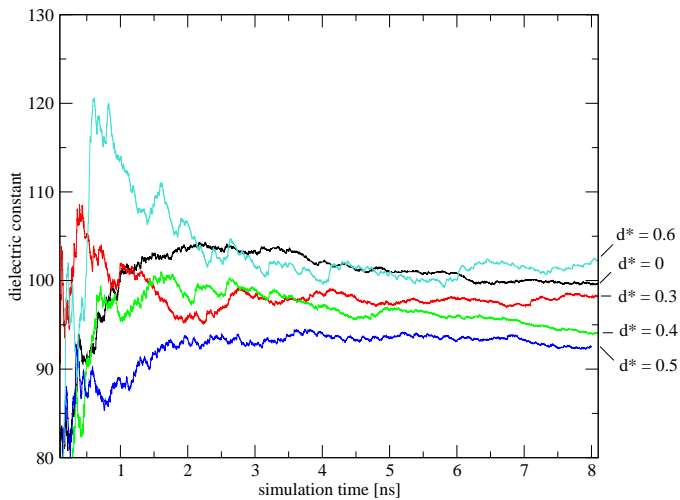}
\caption{\label{fig5}Convergence of $\eps$ with simulation time, for dipole elongations $d^*=d/\sig=0$, $0.3$, $0.4$, $0.5$ and $0.6$.}
\end{center}
\end{figure}

Long runs were needed to obtain even moderate accuracy (about $2\%$) in the estimated dielectric constants.  Fig.~\ref{fig5} shows the running estimate of $\eps$ as a function of simulation time. The slow convergence, especially for large elongations of the dipole, can be traced back to the long relaxation times $\tau_M$, as shown by the following error analysis.

By definition, the probability distribution of the sample having a total dipole moment of magnitude $M$ and arbitrary orientation is given by
\eq{	\label{P(M)}
   P(M) \propto 4\pi M^2 e^{-\be F(M)},
}
where $F(M)$ is the free energy of the system. From macroscopic electrostatics, the energy of a spherical dielectric sample, of dielectric constant $\eps$ and carrying a uniform polarization $M/V$, is
\eq{	\label{U(M)}
    U(M) = \frac{2\pi M^2}{V} \frac{2\eps'+\eps}{(\eps-1)(2\eps'+1)},
}
where $\eps'$ is the dielectric constant of the surrounding medium. Using $F(M)\simeq F(0)+U(M)$, we find from \eqref{P(M)} and \eqref{U(M)} that the fluctuations of  $g=M^2/(N \mu^2)$ are distributed according to
\eq{		\label{P(g)}
   P(g) = A \sqrt{g}\, {e}^{- \kappa g}, \quad
   \kappa \equiv \frac{9y}{2} \frac{2\eps'+\eps}{(\eps-1)(2\eps'+1)},
}
where the normalization constant is $A=2\sqrt{{\kappa^3}/{\pi}}$. The mean of this distribution is $\moy{g}=3/(2\kappa)$, in agreement with the fluctuation formula~\eqref{KF}.
Though the distribution~\eqref{P(g)} is approximate, since we neglected changes in entropy and eq.~\eqref{U(M)} is valid only in the linear regime, it gives a good description of fluctuations of the total dipole moment observed in simulations of polar fluids~\cite{KusManSvi}.

The dielectric relaxation time $\tau_M$ gives a time scale for two measurements of ${M}^2$ to be independent. In a simulation of total duration $t$, the distribution \eqref{P(g)} is thus sampled $n\simeq t/\tau_M$ times. After $n$ such independent measurements, the standard deviation in the average $\sum_{i=1}^n g_i/n$ of the $g$ factor is $\sig_{g,\,n} = \sig_g/\sqrt{n}$ where $\sig_g^2=\moy{(g-\moy{g})^2}=3/(2\kappa^2)$ is the variance of the distribution \eqref{P(g)}. The expected relative uncertainty in the $g$ factor is therefore
\eq{
    I_{\moy{g}} \equiv \frac{\sig_{g,\,n}}{\moy{g}} = \sqrt{\frac{2}{3n}} = \sqrt{\frac{2}{3}\frac{\tau_M}{t}}.
}
Solving \eqref{KF} for $\eps$, one has
\eq{
   \eps - 1 = \frac{3y\moy{g}(2\eps'+1)}{2\eps'+1-3y\moy{g}}.
}
By the rules of propagation of errors, the relative uncertainty in the dielectric constant minus one is thus
\eq{	\label{I_eps}
   I_{\eps-1} = \frac{2\eps'+\eps}{2\eps'+1} \sqrt{\frac{2}{3}\frac{\tau_M}{t}}.
}
The error bars in Fig.~\ref{fig3} were determined from this formula, and are in agreement with the fluctuations observed in Fig.~\ref{fig5}. In a Debye dielectric, the relaxation time $\tau_M$ is related to the Debye relaxation time $\tau_D$ (which is independent of boundary conditions) by~\cite{NeuSte}
\eq{	\label{tau_D}
   \tau_M = \frac{2\eps'+1}{2\eps'+\eps} \tau_D.
}
Inserting \eqref{tau_D} into \eqref{I_eps}, we see that larger values of~$\eps'$ will lead to smaller uncertainties in the dielectric constant. This explains the faster convergence of $\eps$ observed when using metallic boundary conditions~\cite{deLPerSmi,WanHolMul}.

According to the present analysis, the slow convergence of $\eps$, as determined from the fluctuation formula, is due to the large value of the Debye dielectric relaxation time~\cite{KurMarCha} and the rather broad distribution $P(g)$. Moreover, the uncertainties in $\eps$ are independent of system size, as long as it is macroscopic. In large systems, it may therefore be more favourable to determine $\eps$ from correlation functions rather than from the fluctuation formula.

\section{Structure}
\label{S5}

\subsection{The pair distribution function}

The pair distribution function $h(1,2)=h(\ver,\vmu_1,\vmu_2)$ of the infinite system can be expanded in rotational invariants~\cite{GraGub}:
\eq{	\notag
   h(1,2) = h^{000}(r) + h^{110}(r)\Phi^{110}(1,2) + h^{112}(r)\Phi^{112}(1,2) + \ldots
}
where
\begin{gather}
\Phi^{110}(1,2)=\hat{\vect{\mu}}_1\cdot\hat{\vect{\mu}_2}
\\
\Phi^{112}(1,2) = 3(\hat{\vect{\mu}}_1\cdot\hat{\ver})(\hat{\vect{\mu}}_2\cdot\hat{\ver}) - \hat{\vect{\mu}}_1\cdot\hat{\vect{\mu}}_2.
\end{gather}
The $\Phi^{l_1l_2l}$'s form on orthogonal basis for the angular dependence of $h(1,2)$. The first projections are
\begin{gather}	\label{h000}
h^{000}(r) = \moy{h(1,2)}_{\vmu_1,\vmu_2} =  g(r)-1
\\	\label{h110}
h^{110}(r) = 3 \moy{h(1,2) \Phi^{110}(1,2)}_{\vmu_1,\vmu_2}
\\	\label{h112}
h^{112}(r) = \frac{3}{2} \moy{h(1,2) \Phi^{112}(1,2)}_{\vmu_1,\vmu_2}
\end{gather}
where $\moy{\cdots}_{\vmu}=\int\cdots\romd\Omega_{\vmu}/4\pi$ denotes an unweighted angular average over the orientations of $\vmu$.

\begin{figure}[h]
\begin{center}
\epsfig{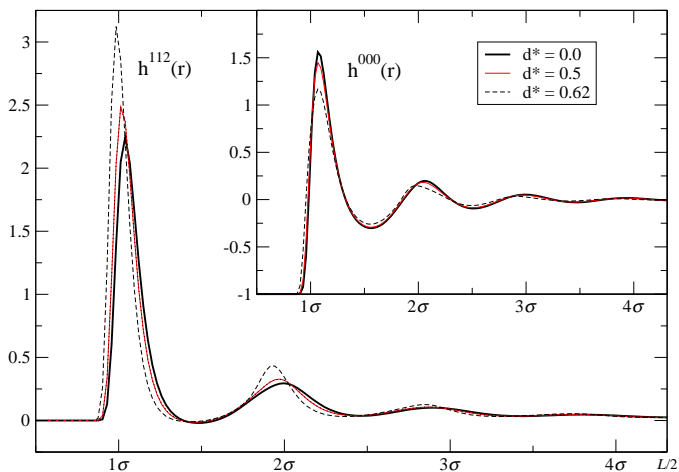}
\caption{\label{fig6}Projections $h^{000}(r)$ and $h^{112}(r)$ of the pair correlation function for three values of $d^*=d/\sig$.}
\end{center}
\end{figure}

Plots of $h^{000}(r)$ and $h^{112}(r)$ are shown on Fig.~\ref{fig6} for three elongations $d$ of the dipole. As $d$ is increased, the stronger multipolar moments carried by the molecules lead to a slight reduction of the fluid structure as measured by the centre-to-centre distribution $g(r)$, but more orientational order, as measured by the projections $h^{112}(r)$ and $h^{110}(r)$ (the latter projection, not shown in the figure, closely resembles $h^{112}(r)$).

The projection $h^{112}(r)$ is related to the dielectric constant of the fluid by the formula
\eq{	\label{eps ND}
   \lim_{r\to\infty} r^3 h^{112}(r) = \frac{(\eps-1)^2}{\eps} \frac{1}{4\pi\rho y}
}
first derived by Nienhuis and Deutch~\cite{NieDeu}. A 512-molecule system with a half box size of $L/2\sig=4.3$ is too small to reach the asymptotic limit~\eqref{eps ND}. The results for the correlation function shown below in Figs.~\ref{fig7}--\ref{fig10} were hence obtained using a larger system (simulation of $5555$ molecules during $6$\,ns) under the same conditions ($\rho^*=0.8$, $T^*=1.35$, $\mu^*=2$, $d=\sig/2$, $\eps'=\infty$). Now $L/2\sig=9.55$, and Fig.~\ref{fig7} shows that $r^3h^{112}(r)$ does reach the asymptotic value~\eqref{eps ND} at a distance $r\simeq 7\sig$, as in the case of point dipoles~\cite{Kus91}. Contrary to the fluctuation formula, eq.~\eqref{eps ND} gives estimates of $\eps$ that become more accurate when the size of the system is increased.

Kusalik made the observation that $r^3h^{112}(r)$ drops sharply for $r$ greater than $L/2$, even when the reduced size of the volume element is properly taken into account in the normalisation \cite{Kus90,Kus91}. The reason for this sharp drop is simple: the Ewald potential deviates strongly from the Coulomb potential at distances greater than $L/2$ (it vanishes in particular at a finite distance between $L/2$ and $\sqrt{3}L/2$ depending on the position of the charges). The rapid decrease in $r^3h^{112}(r)$ reflects thus an artificial finite size effect arising from the Ewald summations of the periodically replicated system~\footnote{In the reaction-field method, the correlation functions are already unreliable at distances of the order of the cut-off distance.}. In a system with long-range forces, great care must therefore be exercised in the interpretation of correlation functions at distance larger than half the box length.

\begin{figure}[h]
\begin{center}
\psfrag{r3}{$r^3 h^{112}(r)$}
\epsfig{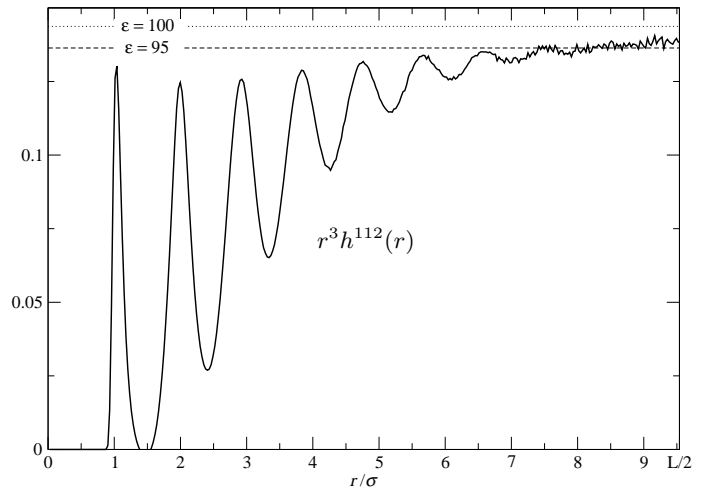}
\caption{\label{fig7}Convergence of $r^3h^{112}(r)$ at large distances towards the limit \eqref{eps ND}. Data from a 6\,ns long simulation of a system of 5555 molecules ($\rho^*=0.8$, $T^*=1.35$, $\mu^*=2$, $d^*=0.5$).}
\end{center}
\end{figure}

\begin{figure}[h]
\begin{center}
\psfrag{r2}{$r^2 (h^{110}(r)-h^{110}(\infty))$}
\epsfig{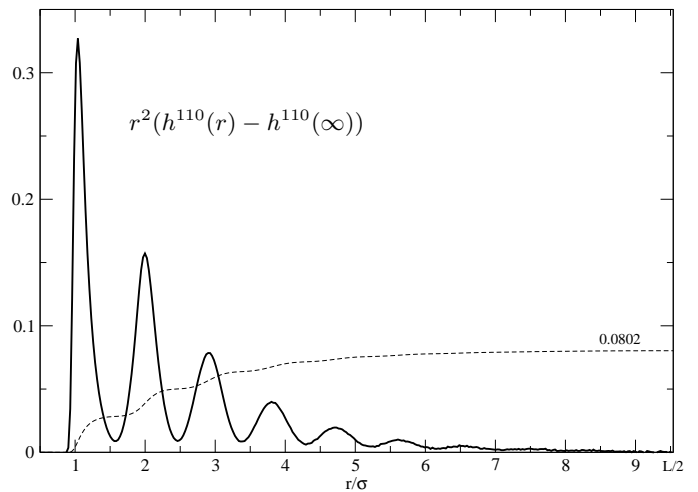}
\caption{\label{fig8}The function $r^2 (h^{110}(r)-h^{110}(\infty))$ and its integral (dashed line). Same system as in Fig.~\ref{fig7}.}
\end{center}
\end{figure}

The projection $h^{110}(r)$ is also related to the dielectric constant, since the fluctuation formula~\eqref{KF} can be written in the Kirkwood form~\footnote{For a cubic simulation cell, the upper limit in the integral becomes $\sqrt{3}L/2$.}
\eq{	\label{KF corr}
   \frac{(\eps - 1)(2\eps'+1)}{2\eps'+\eps} = 3y \left(1 + \frac{4\pi\rho}{3}\int_0^{\infty}\!h_{\eps'}^{110}(r) r^2 \romd r \right).
}
Since the l.h.s. of eq.~\eqref{KF corr} depends on $\eps'$, the projection $h^{110}(r)$ must also be sensitive to this boundary condition, whence the introduction of a subscript $\eps'$. Fig.~\ref{fig8} shows a plot of $r^2 h_{\infty}^{110}(r)$, and the integral of this function, for the same system as in Fig.~\ref{fig7}. The correlations extend up to $r\simeq7\sig$, just as in the case of $h^{112}(r)$. This distance corresponds to the scale beyond which the fluid behaves as a continuum dielectric and obeys the equations of macroscopic electrostatics.

Though formula \eqref{KF corr} is equivalent to \eqref{KF}, it is worthwhile to discuss how the pair correlation function depends on the dielectric constant of the external medium. This problem has been addressed in Ref.~\cite{NieDeu} (see also the perturbation theory presented in Ref.~\cite{deLPerSmi}); here, we hope to give a clear and concise answer to the above question, using simple physical arguments to justify the formulae.

In a spherical sample, $h^{110}_{\eps'}(r)$ is in fact the only projection, among all $h^{l_1l_2l}$'s, to be strongly affected by the boundary condition $\eps'$. This is due to the surface term in the Hamiltonian~\eqref{H}, which corresponds to an interaction energy between two molecules
\eq{	\label{int sph}
   \frac{4\pi}{2\eps'+1} \frac{\vect{\mu}_1\cdot\vect{\mu}_2}{V}
}
which has the angular dependence of $\Phi^{110}(1,2)$.

As the interaction \eqref{int sph} is independent of the distance between the molecules, $h_{\eps'}^{110}(r)$ does not decay in general to zero at infinity, but rather to an $\eps'$-dependent constant. We will prove below that this constant is
\eq{	\label{asympt h110}
   \lim_{r\to\infty} h_{\eps'}^{110}(r) = \frac{2}{V} \frac{(\eps-1)^2}{3\rho y\eps} \frac{\eps'-\eps}{2\eps'+\eps}
}
in the spherical geometry.
The constant \eqref{asympt h110} vanishes only when using the boundary condition $\eps'=\eps$, which mimicks an infinite sample, or in the thermodynamic limit $V\to\infty$ (which is never reached in simulations).

The fact that $h_{\eps'}^{110}(r)$ contains the ${\cal O}(1/V)$ constant contribution~\eqref{asympt h110} at large distances ensures that the Kirkwood formula~\eqref{KF corr} gives consistent results for different boundary conditions. Indeed, when $h_{\eps'}^{110}(r)$ is integrated over the volume $V=4\pi R^3/3$ of a large sample, as in the r.h.s of eq.~\eqref{KF corr}, the constant \eqref{asympt h110} gives a finite contribution to the integral that is precisely what is required to match the $\eps'$-dependence of the l.h.s. of the equation. In other words, eqs. \eqref{int sph} and \eqref{asympt h110} imply that
\begin{equation}
\notag
   4\pi\int_0^R h_{\eps'}^{110}(r)r^2\romd r =  4\pi \int_0^R h_{\eps}^{110}(r)r^2\romd r + V  h_{\eps'}^{110}(\infty),
\end{equation}
when the samples are large enough ({\it i.e.} in the limit $R\to\infty$). When this identity is inserted into the Kirkwood formula~\eqref{KF corr}, it is immediately clear that the predicted dielectric constant is independent of~$\eps'$.

In order to prove eq.~\eqref{asympt h110} with simple physical arguments, we need to recall two basic results from the statistical mechanics of polar liquids. The first result is the expression of the density $\rho(\ver,\vmu)$ of molecules at $\ver$ with orientation~$\vmu$ in a polarised sample permeated by a macroscopic field $\vect{E}(\ver)$: \cite{AlaBal}
\eq{	\label{rho(1)}
   \rho(\ver,\vmu) = \frac{\rho}{4\pi} \left(1+\frac{\eps-1}{3y}\beta \vmu\cdot\vect{E}(\ver)\right) + {\cal O}(E^2)
}
($y$ is defined after eq.~\eqref{KF cond}). This formula is consistent with the constitutive relation $\vect{P}(\ver)=(\eps-1)/(4\pi)\vect{E}(\ver)$ of macroscopic electrostatics, since the average polarisation density in the fluid is by definition $\vect{P}(\ver)=\int\!\rho(\ver,\vmu)\vmu\,\romd\Omega_{\vmu}$.
 
The second result we need is the expression of the effective dipole moment $\vmu^{\rm eff}$ of a polar molecule held fixed in a polar liquid ($\vmu^{\rm eff}$ is defined as $\vmu$, the dipole moment of the fixed molecule, plus the total dipole moment of the screening cloud around~$\vmu)$. One may first think naively that $\vmu^{\rm eff}=\vmu/\eps$: the fluid would screen the dipole according to its dielectric constant. But this would be treating the polar fluid as a dielectric continuum everywhere, including in the interior of the fixed molecule, which is obviously wrong. An exact statistical mechanical calculation shows that the right answer is~\cite{FinBalHan}
\eq{	\label{mu eff}
   \vmu^{\rm eff} = \frac{\eps-1}{3 y \eps} \vmu.
}
(The expression $(\eps-1)/3y\eps$ can itself be interpreted as being composed of a factor $(\eps-1)/3y$ arising from local correlations around~$\vmu$, times the expected factor $1/\eps$ due to screening by distant molecules). With these two results in mind, we can now understand easily formulae \eqref{eps ND}, \eqref{int sph} and \eqref{asympt h110}.

The result \eqref{eps ND} for the large distance behaviour of the pair correlation function $h(1,2)$ can be seen as a straightforward consequence of the screening effect~\eqref{mu eff}. By definition of the distribution functions, the density of molecules at $\ver_2$ with orientation $\vmu_2$, when a molecule is known to be located at $1=(\ver_1,\vmu_1)$ is
\eq{	\label{rho(1|2)}
   \rho(2|1) = \frac{\rho(2,1)}{\rho(1)} = \frac{\rho}{4\pi}(1+h(1,2)).
}
From~\eqref{mu eff}, the electric field due to the fixed molecule $\vmu_1$ is equivalent, at large distances $\ver_{12}=\ver_2-\ver_1$, to that of a renormalized dipole moment~$\vmu_1^{\rm eff}$. This dipolar field $-\grad_2(\vmu_1^{\rm eff}\cdot\grad_1)1/|\ver_{12}|$ is locally uniform and weak, so we can apply the linear response result~\eqref{rho(1)}. Using \eqref{rho(1)} and \eqref{mu eff}, we find that
\eq{	\label{rho(1|2) as}
   \rho(2|1) \mathop{\sim}\limits_{|\ver_{12}|\to\infty} \frac{\rho}{4\pi}\left(1-\frac{(\eps-1)^2}{9y^2\eps}\beta v_{\rm dip}(1,2)\right)
}
where $v_{\rm dip}(1,2)=(\vmu_1\cdot\grad_1)(\vmu_2\cdot\grad_2){1}/{|\ver_{12}|}$ is the dipolar potential, and we assume an infinitely extended sample. Comparing \eqref{rho(1|2)} and \eqref{rho(1|2) as}, we obtain the asymptotic behaviour of the pair correlation function:
\eq{	\label{h(1,2) as}
   h(1,2) \sim \frac{(\eps-1)^2}{9y^2\eps} (-\beta v_{\rm dip}(1,2)), \qquad |\ver_{12}|\to\infty.
}
The result \eqref{eps ND} follows then upon inserting \eqref{h(1,2) as} into~\eqref{h112}.

Formula~\eqref{asympt h110} can be interpreted in a similar way, namely as arising from the interaction of a molecule with the reaction field produced by the screened dipole moment of another molecule. We recall from macroscopic electrostatics that a dipole $\vmu_1$ at the centre of a spherical sample of radius~$R$ and dielectric constant $\eps$, surrounded by a dielectric medium~$\eps'$, produces a uniform reaction field
\eq{	\label{E_R cont}
   \vect{E}^{[\eps,\eps']}_R(\vmu_1) = \frac{2}{\eps}\frac{\eps'-\eps}{2\eps'+\eps}\frac{\vmu_1}{R^3}
}
inside the sample, because of the surface charge density induced at the dielectric discontinuity~\cite{Fro}. In a finite spherical sample, a molecule at a position $\ver_2$ far enough from $\vmu_1$ -- so that it does not disturb the screening cloud around it -- will interact therefore not only with the dipolar field of~$\vmu_1^{\rm eff}$, as in~\eqref{rho(1|2) as}, but also with the reaction field $\vect{E}_R(\vmu_1^{\rm eff})$ produced by the screened dipole moment of this molecule. From~\eqref{rho(1)}, a term
\eq{	\label{int E_R}
   \frac{\rho}{4\pi}\cdot\frac{\eps-1}{3y}\beta \vmu_2\cdot{\vect{E}}^{[\eps,\eps']}_R(\vmu_1^{\rm eff})
}
must hence be added to \eqref{rho(1|2) as} in a finite spherical sample. The pair correlation at large distances, eq.~\eqref{h(1,2) as}, includes then the additional contribution
\eq{	\label{prec}
   \frac{4\pi}{3V}\frac{2(\eps-1)^2}{9y^2\eps}\frac{\eps'-\eps}{2\eps'+\eps} \,\beta \vmu_2\cdot\vmu_1
}
where we used \eqref{E_R cont}, \eqref{mu eff} and $V=4\pi R^3/3$. Formula \eqref{asympt h110} follows now from projecting \eqref{prec} onto $\Phi^{110}(1,2)=\hat{\vmu}_1\cdot\hat{\vmu}_2$ according to~\eqref{h110}.

We conclude this discussion by noting that the interaction energy~\eqref{int sph} used in the simulations, or equivalently the surface term in the Hamiltonian~\eqref{H}, can also be interpreted in terms of a reaction field effect. Indeed, each polar molecule in the sample will create a reaction field, acting on itself and on all other molecules, that is given by eq.~\eqref{E_R cont} with $\eps=1$. Since the Ewald sums \eqref{def Psi} give the electrostatic energy between the molecules in the case of a sample surrounded by a metal ($\eps'=\infty$), the correction to this energy to be used in the Hamiltonian of a spherical system with boundary condition~$\eps'$ is
\eq{
   \frac{1}{2}\sum_{i,j=1}^N (-\vmu_i)\cdot\left[(\vect{E}_R^{[1,\eps']}(\vmu_j)-\vect{E}_R^{[1,\infty]}(\vmu_j)\right] = 
   \frac{2\pi \vect{M}^2}{\eps'+1},
}
in agreement with \eqref{H}.

\subsection{Site-site correlations}

Contrary to the point dipolar fluid model, the present model with extended dipoles has well defined site-site distribution functions $h_{++}(r)=h_{--}(r)$ and $h_{+-}(r)$ \cite{HanMcD}. From these, we get the charge-charge correlation function $S(r)=S_{\rm intra}(r) + S_{\rm inter}(r)$, where
\eq{
   S_{\rm inter}(r) = 2 q^2 \rho^2 (h_{++}(r) - h_{+-}(r)),
}
describes the inter-molecular correlations, while
\eq{
   S_{\rm intra}(r) = 2 q^2 \rho \delta(\ver) - 2 q^2 \rho \frac{\delta(|\ver|-d)}{4\pi d^2}
}
is the intra-molecular correlation function for a molecule with a rigid dipole of extension~$d$. The charge-charge correlation is of special interest, because it satisfies the two sum rules~\cite{Mar}:
\begin{gather}
\hspace{-3mm}\text{Neutrality:}\quad \int\! S(r)\,\romd^3\ver = 0\\	\label{SL}
\hspace{-3mm}\text{Stillinger-Lovett:}\quad \frac{1}{\eps} = 1+\frac{2 \pi \beta}{3}  \int\romd^3\ver\, \ver^2\, S(r).
\end{gather}

\begin{figure}[h]
\begin{center}
\psfrag{S(r)}{$h_{++}(r)-h_{+-}(r)\sim S(r)$}
\psfrag{h++}{$h_{++}(r)$}
\psfrag{h+-}{$h_{+-}(r)$}
\epsfig{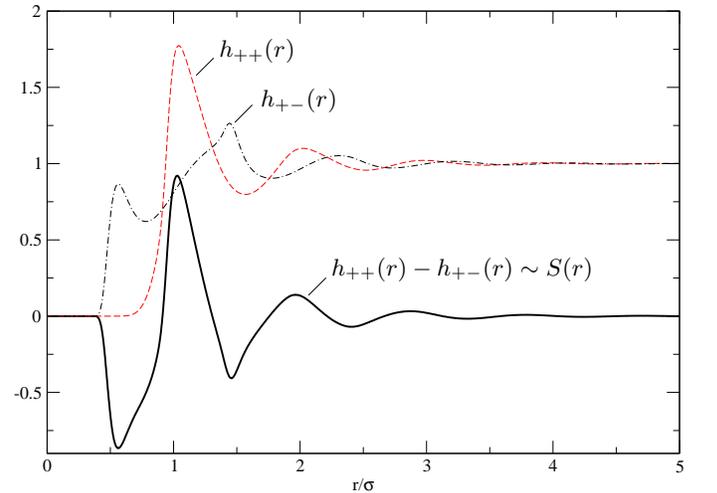}
\caption{\label{fig9}Site-site distribution functions. Same system as in Fig.~\ref{fig7}.}
\end{center}
\end{figure}

\begin{figure}[h]
\begin{center}
\psfrag{int}{$4\pi\int_0^r S_{\rm inter}(r')r'^4\romd r'$}
\psfrag{limit}{$(1-\frac{\eps-1}{3y\eps})\frac{d^2}{\rho}$}
\epsfig{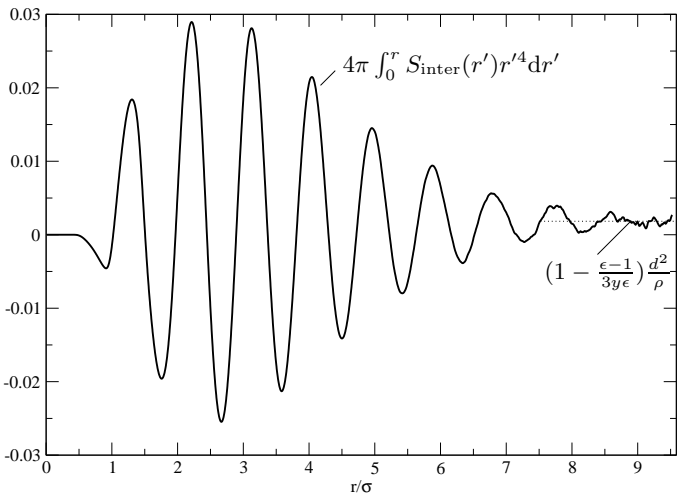}
\caption{\label{fig10}Integration of the second moment of $S_{\rm inter}(r)$.}
\end{center}
\end{figure}

The site-site correlation functions and $S(r)$ are shown on Fig.~\ref{fig9} for $d=\sig/2$.
The charge neutrality sum rule is found to be accurately satisfied:
\eq{
   \rho \int_0^{\infty} (h_{++}(r) - h_{+-}(r)) r^2 \romd r \simeq 8.8\cdot10^{-5}.
}
The Stillinger-Lovett sum rule allows in principle the determination of the dielectric constant from~$S(r)$, but this route is not practicable in a computer simulation, because of the unfavourable ratio $(1-\eps)/\eps$ which saturates for large~$\eps$, and also because it is difficult to determine the second moment of~$S_{\rm inter}(r)$ accurately. Fig.~\ref{fig10} shows however that eq.~\eqref{SL} is satisfied within the uncertainties of our data.

\section{Orientational order}
\label{S6}

When the molecular dipole has an extension greater than $d\simeq0.64\sig$, the simulations show spontaneous formation of orientationaly ordered phases, starting from random initial configurations. At the state point under consideration ($\rho^*=0.8$, $T^*=1.35$), we observed phase coexistence between a dense liquid crystal and a very dilute gas. In order to deal with pure phases, we performed simulations at constant pressure ($p^*=p\,\sig^3/u=0.22$), rather than constant volume, for $d\geq 0.62\sig$.

The occurrence of orientational order was monitored by computing two order parameters. The rank-one order parameter $P_1$ is defined as
\eq{
   P_1 = \frac{\moy{{M}_{\parallel}}}{N \mu},
}
where ${M}_{\parallel}\equiv\vect{M}\cdot \hat{\vect{n}}$ is the projection of the total dipole moment along the director~$\vect{n}$ ($P_1=1$ for a completely polarized system). The second-rank order parameter $P_2$ is the largest eigenvalue of the matrix
\eq{
   Q_{\al\be} = \frac{1}{N \mu^2} \moy{\sum_{i=1}^N \frac{1}{2}(3 {\mu}_i^\al {\mu}_i^\be - \mu^2 \delta_{\al\be})}
}
where ${\mu}^\al_i$ is the $\al$ component of the vector $\vect{\mu}_i$. The corresponding eigenvector~$\vect{n}$ is the director. $P_2 = 1$ when all dipoles are oriented parallel to $\vect{n}$ or~$-\vect{n}$.


Table~\ref{T2} lists our results for these order parameters, as well as for the dielectric tensor $\boldsymbol{\eps}=(\eps_{\parallel},\eps_{\perp})$. In a liquid crystal with director~$\vect{n}$, the latter is determined by the following generalization of eq.~\eqref{KF cond}: 
\begin{equation}
 \eps_{\parallel} = 1 + y\, \frac{{\big\langle{M}_{\parallel}^2\big\rangle}-\moy{{M}_{\parallel}}^2}{N\mu^2},
\end{equation}
and a similar equation for $\eps_{\perp}$ in terms of the perpendicular fluctuations $\big({\big\langle\vect{M}_{\perp}^2\big\rangle}-\moy{\vect{M}_{\perp}}^2\big)/N\mu^2$.

\hdashlinewidth .5pt
\hdashlinegap 1.25pt

\begin{table}[h]
\begin{center}
\begin{tabular*}{0.65\columnwidth}{@{\extracolsep{\fill}}cccccc}
\hline
$d/\sig$& $\moy{\rho^*}$ &	$P_1$	&	$P_2$	&	$\eps_{\parallel}$	&	$\eps_{\perp}$	\\	\hline
$0.62$	&0.80	&	0.08	& 	0.07 	&	\multicolumn{2}{c}{$\eps=103.8(\pm4)$}	\\
0.63	&0.80	&	0.09	&	0.08 	&	\multicolumn{2}{c}{$\eps=112.6(\pm5)$}		\\ \hdashline 
0.64	&0.95 	&	0.97	&	0.91	&	1.16	&	1.46			\\
0.65	&0.98	&	0.66	&	0.91	&	1.28	&	1.56			\\
0.66	&1.03	&	0.98	&	0.94	&	1.02	&	1.43			\\
\hline
\end{tabular*}
\end{center}
\caption{\small\label{T2} Constant pressure simulations of the dipolar soft-sphere system at $p^*=0.22$, $T^*=1.35$ and $\mu^*=2$.
Data from 8\,ns long simulations of 512 molecules, collected after an equilibration period that lasted up to 10\,ns. For $d\geq 0.64\sig$, the system is a ferro-electric liquid crystal with columnar order. Uncertainties in $\eps_{\parallel}$ and $\eps_{\perp}$ are about $\pm0.01$ and $\pm 0.02$ respectively.}
\end{table}

\begin{figure}
\begin{center}
\epsfig{file=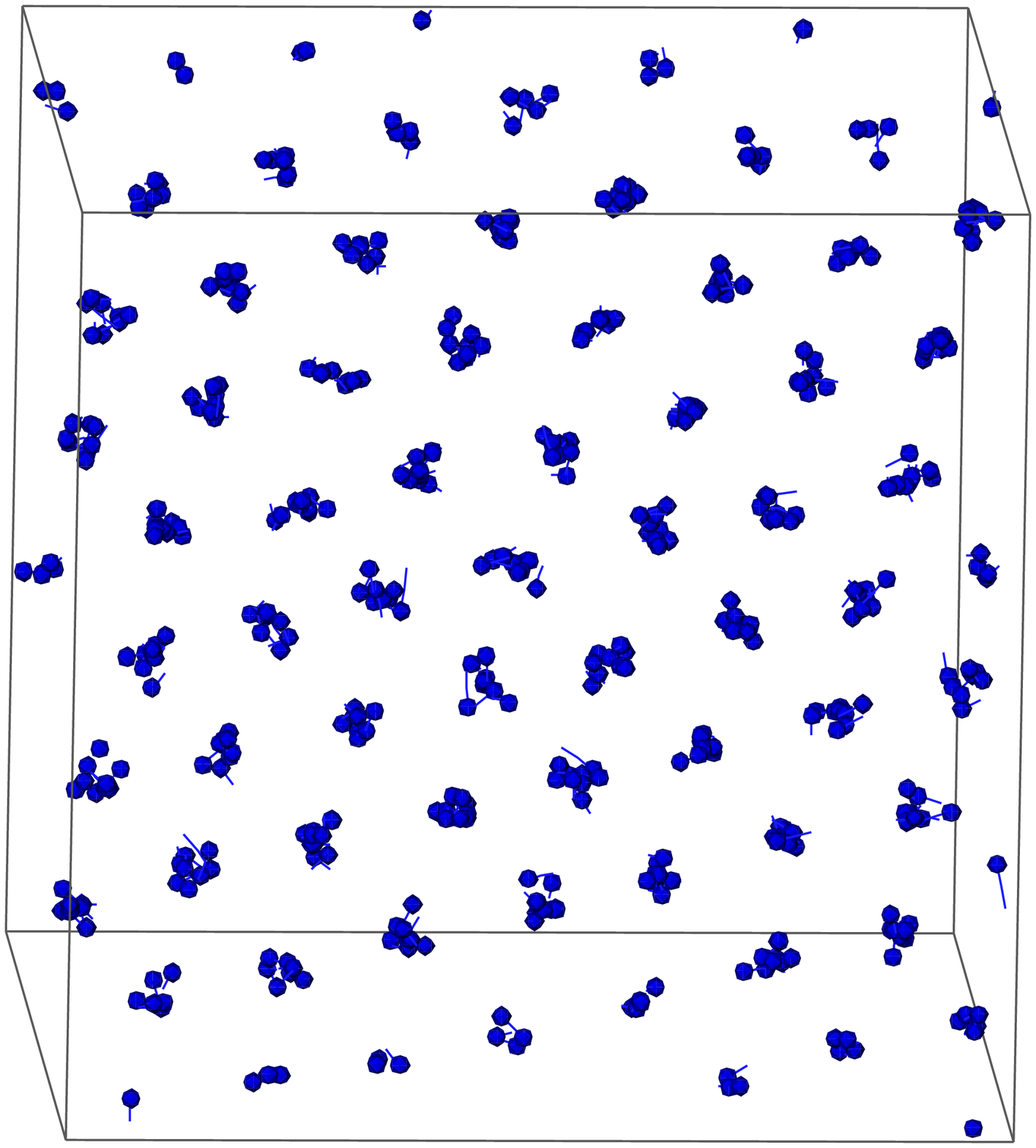,scale=0.2}\epsfig{file=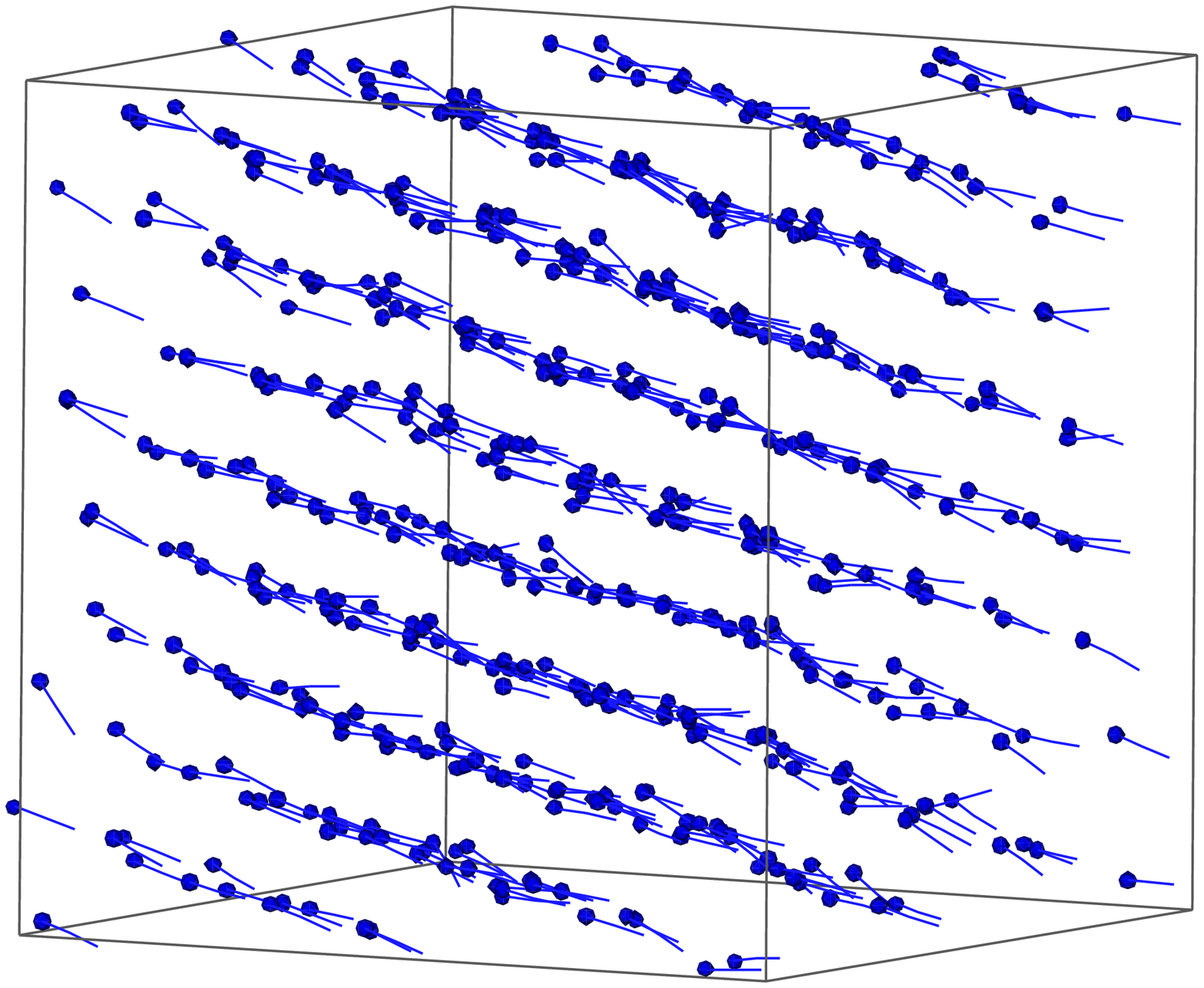,scale=0.25}
\caption{\small\label{fig11} Snapshots of the simulation cell of a dipolar soft-spheres fluid in the columnar phase. The hexagonal lattice in the plane orthogonal to the director is apparent in the first snapshot. The dipoles are represented by a line joining the minus charge (shown as a small bead) to the plus charge.}
\end{center}
\end{figure}

When $d$ is increased above the critical distance $d_c\simeq 0.63\sig$, the order parameter $P_2$ jumps from essentially zero to about almost one, indicating a first order transition to a highly orientationally ordered phase. Fig.~\ref{fig11} provides snapshots of the simulation cell for $d=0.64\sig$. It is clear from the snapshots that the molecules are associated into columns, composed of chains of dipoles oriented head-to-tail. These columns are all parallel to the director, and are arranged in a hexagonal lattice in the perpendicular plane. The simulations for $d>0.64\sig$ yielded similar liquid crystal phases with columnar order, each with a different orientation of the director. The system shows strong spatial correFlations in the direction of the director, but it is still fluid in that direction, as indicated by the mean-square displacement of the molecules. The latter increases indeed linearly with time, with a diffusion constant of about $D_{\parallel}\simeq 0.09\cdot10^{-5}$\,cm$^2$/s.

In some runs (not listed in Table~\ref{T2}), the system formed two liquid crystal domains characterized by different orientations of the director. As the transition to a single domain is expected to occur on a time scale much larger than our simulation time (8\,ns), since it requires the collective motion of many molecules, we included in Table~\ref{T2} only results from runs where a single domain formed spontaneously in less than 10\,ns. Most runs yielded fully polarized liquid crystals ($P_1$ close to~1). It is likely that the lower value of $P_1$ measured in the case $d=0.65\sig$ is due to insufficient sampling of phase space: the probability of a column inverting its orientation during our simulation time is indeed very small.

As the column configuration is energetically more favourable for extended than for point dipoles (see Fig.~\ref{fig2}), it is not surprising that liquid crystal columnar phases form at a much lower dipolar coupling constant than previously reported for point dipolar fluid models; here $\mu^*{}^2/T^*=\mu^2/\sig^3 kT\simeq 3$, while columnar phases were observed in the dipolar soft-sphere fluid at $\mu^*{}^2/T^*=9$, and in the dipolar hard sphere fluid at $\mu^*{}^2/T^*=6.25$ \cite{WeiPat,WeiLevZar}. At the same state point as under consideration here, Kusalik found however that even a strong electric field did not induce a nematic phase in the dipolar soft-sphere model with point dipoles~\cite{Kus94}. The hexagonal lattice arrangement found here is to be contrasted with the square lattice reported in Ref.~\cite{WeiLevZar} (the lattice type was not specified in Ref.~\cite{WeiPat}).


\section{Conclusion}

We have extended the considerable body of earlier work on dipolar fluids by replacing the usual point-dipole on spherical molecules by physically more relevant extended dipoles obtained by placing two opposite charges symmetrically with respect to the centre. The structural, dielectric and dynamical behaviour was monitored as the spacing $d$ of the charges was increased, keeping the total dipole moment $\mu=q d$ fixed. Periodic boundary conditions were used with proper Ewald summations of the Coulombic interactions within an infinitely large sphere bounded by a dielectric medium of permittivity $\eps'$. The key findings are the following:

a) Runs of several nanoseconds, longer than in most previous studies, were required to obtain estimates of the dielectric constant $\eps$ within about 2\% with the standard fluctuation formula~\eqref{KF}. A careful error analysis shows that ``metallic'' boundary conditions (where $\eps'=\infty$ at infinity) yield optimal estimates of~$\eps$.

b) The values of $\eps$ deduced from the $h^{112}$ and $h^{110}$ correlation functions agree with the fluctuation results within statistical errors, provided a sufficiently large simulation cell is used to obtain proper estimates of the asymptotic behaviour of these correlation functions. The strong influence of the boundary condition~$\eps'$ on the projection $h^{110}(r)$ arises from the interaction of the polar molecules with the reaction field due to the dielectric discontinuity between the fluid and the external medium~$\eps'$. When $\eps'\neq \eps$, $h^{110}(r)$ does not decay to zero at large distances, but rather to a finite constant of order $1/V$. We derived the value of this constant [eq.~\eqref{asympt h110}] by using simple physical arguments based on macroscopic electrostatics and linear response theory.


c) $\eps$ has a non-trivial dependence on the ratio $d^*=d/\sig$. Up to $d^*\simeq 0.25$, $\eps$~agrees with the point-dipole result within statistical uncertainties, thus illustrating the practical usefulness of the point dipole limit. For $d^*\gtrsim 0.3$, $\eps$~drops to a minimum value roughly 6\% below the point-dipole result when $d^*\simeq 0.55$. When $d^*$ is further increased, $\eps$ increases sharply and reaches a maximum at $d^*\simeq 0.6$.

d) For still larger extensions $d^*$, the system is seen to undergo a transition, at constant pressure, to an orientationally ordered stated similar to a columnar phase with a hexagonal ordering in the plane orthogonal to the director. This phase is characterized by large values of the usual orientational order parameters $P_1$ and $P_2$. At the same time the dielectric tensor becomes anisotropic, and the mean dielectric constant is very low ($\eps\simeq 1.4$), signalling the strong suppression of dipole moment fluctuations. The transition to the columnar phase occurs at a value of the dipole moment well below that required to observe the transition with point dipoles~\cite{WeiPat,WeiLevZar}.

e) The dynamics, characterized by the relaxation times $\tau_M$ and $\tau_\mu$ of the total and individual dipole moments, as well as by the self diffusion constant $D$, slows down very significantly as $d^*$ increases, due to the enhanced tendency of the system to form parallel strings, which eventually lead to the columnar phase. In the latter, the diffusion coefficient $D_{\parallel}$ parallel to the director is about two order of magnitude smaller than $D$ in the isotropic phase, but still substantial, showing that the system behaves like a one-dimensional fluid.

The present results are for a single high density $\rho^*=0.8$, and a single pressure in the anisotropic phase [$p^*=0.22$, corresponding to $\moy{\rho^*}\simeq 1$]. Clearly more work is needed to be able to map out a complete phase diagram, in view of the additional variable $d^*$. The present work illustrates the strengths and deficiencies of the point-dipole model. Many simple molecular systems fall in a range $d^*\simeq 0.5$, where the deviations from point dipole behaviour begin to be substantial.

\noindent \textbf{Acknowledgements} The authors thank Joachim Dzubiella, Reimar Finken, Michael Klein and Mark Miller for fruitful discussions. The financial support of the Swiss National Science Foundation is also gratefully acknowledged.


\begin{thebibliography}{99}
\bibitem{Ons} L. Onsager, J. Am. Chem. Soc. {\bf 58} (1936): 1486
\bibitem{Kir} J. Kirkwood, {\it The dielectric polarization of polar liquids}, J. Chem. Phys. {\bf 7} (1939): 911-919
\bibitem{MadKiv} For a review, see P.A. Madden and D. Kivelson, {\it A consistent molecular treatment of dielectric phenomena}, Adv. Chem. Phys. {\bf 56} (1984): 467
\bibitem{PatLevWei} G.N. Patey, D. Levesque and J.J. Weis, {\it On the theory and computer simulations of dipolar of dipolar fluids}, Mol. Phys {\bf 45} (1982): 733
\bibitem{FreSmi} See for ex. D. Frenkel and B. Smit, {\it Understanding molecular simulation}, 2nd ed. Academic Press (2002)
\bibitem{deLPerSmi} S.W. de Leeuw, J.W. Perram and E.R. Smith, {\it Simulation of electrostatic systems in periodic boundary conditions. I. Lattice sums and dielectric constants}, Proc. R. Soc. Lond. A {\bf 373} (1980): 27-56; {\it II. Equivalence of boundary conditions}, {\it ibid} {\bf 373} (1980): 57-66; {\it III. Further theory and applications}, {\it ibid} {\bf 388} (1983): 177-193
\bibitem{Neu} M. Neumann, {\it Dipole moment fluctuation formulas in computer simulations of polar systems}, Mol. Phys. {\bf 50} (1983): 841-858
\bibitem{Kus90} P.G. Kusalik, {\it Computer simulation results for the dielectric properties of a highly polar fluid}, J. Chem. Phys. {\bf 93} (1990): 3520-3535
\bibitem{KurMarCha} Z. Kurtovi\'c, M. Marchi and D. Chandler, {\it Umbrella sampling molecular dynamics of the dielectric constant of water}, Mol. Phys. {\bf 78} (1993): 1155-1165
\bibitem{WeiPat} D. Wei and G.N. Patey, {\it Orientational order in simple dipolar liquids: computer simulation of a ferroelectric nematic phase}, Phys. Rev. Lett. {\bf 68} (1992): 2043-2045
\bibitem{WeiLev} J.J. Weis and D. Levesque, {\it Ferroelectric phases in dipolar hard spheres}, Phys. Rev. E {\bf 48} (1993): 3728-3740
\bibitem{GroDie} B. Groh and S. Dietrich, {\it Long-ranged orientational order in dipolar fluids}, Phys. Rev. Lett. {\bf 72} (1994): 2422-2425
\bibitem{KleMcD} M.L. Klein and I.R. McDonald, {\it Structure and dynamics of associated molecular systems. I. Computer simulation of liquid hydrogen fluoride}, J. Chem. Phys. {\bf 71} (1979): 298-308
\bibitem{Smi} M.E. van Leeuwen and B. Smit, {\it What makes a polar liquid a liquid}, Phys. Rev. Lett. 71 (1993): 3991-3994
\bibitem{dipqua} D. Wei, {\it Liquid order in simple polar liquids: effect of the quadrupole moment}, Mol. Cryst. Liq. Cryst. {\bf 269} (1995): 89-98
\bibitem{Cic} G. Ciccotti, M. Ferrario, J.T. Hynes and R. Kapral, {\it Constrained molecular-dynamics and the mean potential for an ion-pair in a polar solvent}, Chem. Phys. {\bf 129} (1989): 241-251
\bibitem{DieMor} A.W. Hertzner, M. Schoen and H. Morgner, {\it The influence of ong range electrostatic forces on static properties of a quasi-Stockmayer fluid}, Mol. Phys. {\bf 73} (1991): 1011-1029
\bibitem{Jac} J.D. Jackson, {\it Classical Electrodynamics}, 3rd ed., John Wiley \& sons, New York (1998)
\bibitem{Kus91} P.G. Kusalik, {\it On the computer simulation of highly polar fluids using large systems}, Mol. Phys. {\bf 73} (1991): 1349-1363
\bibitem{SteGre} M.J. Stevens and G.S. Grest, {\it Structure of soft-sphere dipolar fluids}, Phys. Rev.~E {\bf 51} (1995): 5962-5975
\bibitem{PME} U. Essman, L. Perela, M. L. Berkowitz, T. Darden, H. Lee and L. G. Pedersen, {\it A smooth particle mesh Ewald method},
J. Chem. Phys. {\bf 103} (1995): 8577-8592
\bibitem{gromacs} E. Lindahl, B. Hess and D. van der Spoel, {\it GROMACS 3.0: A package for molecular simulation and trajectory analysis}, J. Mol. Mod. {\bf 7} (2001): 306-317 (http://www.gromacs.org)
\bibitem{WanHolMul} Z. Wang, C. Holm and H.W. M\"uller, {\it Boundary condition effects in the simulation study of equilibrium properties of magnetic dipolar fluids}, J. Chem. Phys. {\bf 119} (2003): 379-378
\bibitem{HocBorBitSte} P. H\"ochtl, S. Boresch, W. Bitomsky and O. Steinhauser, {\it Rationalization of the dielectric properties of common three-site water models in terms of their force field parameters}, J. Chem. Phys. {\bf 109} (1998): 4927-4937
\bibitem{KusManSvi} P.G. Kusalik, M.E. Mandy and I.M. Svishchev, {\it The dielectric constant of polar fluids and the distribution of the total dipole moment}, J. Chem. Phys. {\bf 100} (1994): 7654-7664.
\bibitem{NeuSte} M. Neumann and O. Steinhauser, {\it On the calculation of the frequency-dependent dielectric constant in computer simulations}, Chem. Phys. Lett. {\bf 102} (1983): 508-512
\bibitem{NieDeu} G. Nienhuis and J.M. Deutch, {\it Structure of Dielectric Fluids: I. The two-particle distribution function of polar fluids}, J. Chem. Phys. {\bf 55} (1971): 4213-4229
\bibitem{GraGub} C.G. Gray and K.E. Gubbins, {\it Theory of molecular fluids}, Vol. 1, Clarendon Press, Oxford (1984)
\bibitem{AlaBal} A. Alastuey and V. Ballenegger, {\it Statistical mechanics of dipolar fluids: dielectric constant and sample shape}, Physica A {\bf 279} (2000): 268-286
\bibitem{FinBalHan} R. Finken, V. Ballenegger and J.-P. Hansen, {\it Onsager model for a variable permittivity near an interface}, Mol. Phys. {\bf 101} (2003): 2559-2568
\bibitem{Fro} H. Fr\"ohlich, {\it Theory of dielectrics}, Oxford University Press (1949)
\bibitem{HanMcD} J.-P. Hansen and I.R. McDonald, {\it Theory of simple liquids}, 2nd ed., Academic Press (1986)
\bibitem{Mar} Ph. A. Martin, {\it Sum rules in charged fluids}, Rev. Mod. Phys. {\bf 60} (1988): 1075-1127
\bibitem{WeiLevZar} J.J. Weis, D. Levesque and G.J. Zarragoicoechea, {\it Orientational order in simple dipolar liquid-crystal models}, Phys. Rev. Lett. {\bf 69} (1992): 913-916
\bibitem{Kus94} P.G. Kusalik, {\it Computer simulation study of a highly polar fluid under the influence of static electric fields}, Mol. Phys. {\bf 81} (1994): 199-216
\end{thebibliography}
\end{document}